\pdfoutput=1
\RequirePackage{ifpdf}
\ifpdf % We are running pdfTeX in pdf mode
\documentclass[pdftex]{sigma}
\else
\documentclass{sigma}
\fi

\numberwithin{equation}{section}
\numberwithin{theorem}{section}
\numberwithin{lemma}{section}
\numberwithin{proposition}{section}
\numberwithin{remark}{section}

\begin{document}

\allowdisplaybreaks

\renewcommand{\thefootnote}{$\star$}

\renewcommand{\PaperNumber}{023}

\FirstPageHeading

\ShortArticleName{Dispersionless BKP Hierarchy and Quadrant L\"owner Equation}

\ArticleName{Dispersionless BKP Hierarchy\\
and Quadrant L\"owner Equation\footnote{This paper is a~contribution to the Special Issue in honor of Anatol Kirillov
and Tetsuji Miwa. The full collection is available
at \href{http://www.emis.de/journals/SIGMA/InfiniteAnalysis2013.html}{http://www.emis.de/journals/SIGMA/InfiniteAnalysis2013.html}}}

\Author{Takashi TAKEBE}

\AuthorNameForHeading{T.~Takebe}
\Address{Faculty of Mathematics, National Research University~-- Higher School of Economics,\\
7 Vavilova Str., Moscow, 117312 Russia}
\Email{\href{mailto:ttakebe@hse.ru}{ttakebe@hse.ru}}

\ArticleDates{Received August 23, 2013, in f\/inal form March 10, 2014; Published online March 14, 2014}

\Abstract{We show that $N$-variable reduction of the dispersionless BKP hierarchy is descri\-bed by a~L\"owner type
equation for the quadrant.}


\Classification{37K10; 37K20; 30C55}

\renewcommand{\thefootnote}{\arabic{footnote}}
\setcounter{footnote}{0}

\vspace{-2mm}

\section{Introduction}
\label{sec:intro}

Since an unexpected connection of the dispersionless KP hierarchy (the Benney equations) to the chordal L\"owner
equation was found in the seminal paper~\cite{gib-tsa:99}, similar examples have been found by~\cite{man:04,m-ma-m:02, tak-tak:08,t-t-z:06} and others.
The dispersionless integrable hierarchies are obtained as quasi-classical limits of ``dispersionful'' integrable
hierarchies and thus the Lax operators of the latter are replaced by Lax ``functions''
 (see, for example,~\cite{tak-tak:95}).

On the other hand, L\"owner type equations are dif\/ferential equations characterising one-parameter families of conformal
mappings between families of domains with growing slits and f\/ixed reference domains (the upper half plane for the
chordal L\"owner equation and the unit disk or its exterior for the radial L\"owner equation).
For details of the L\"owner equations in the context of complex analysis we refer to~\cite{a-b-c-d:10, dur:83, pom:75}.

The key point of the ``unexpected'' relation is that {\em the Lax functions satisfying the one-variable reduction
condition are solutions of the L\"owner type equations}.
In order to explain more precisely, let us recall the f\/irst and most typical example found in~\cite{gib-tsa:99}: the
relation between the dispersionless KP hierarchy and the chordal L\"owner equation.
We follow formulation in~\cite{t-t-z:06}.
Suppose that $\mathcal{L}_\mathrm{KP}(\tilde w;t)$ is a~solution of the dispersionless KP hierarchy, which depends on
$t=(t_1, t_2, t_3,\dots)$ only through a~single variable $\lambda$.
Namely, there exists a~function $f_\mathrm{KP}(\tilde w;\lambda) = \sum\limits_{n=0}^\infty u_n(\lambda) \tilde w^{1-n}$
($u_0=1$, $u_1=0$) of two variables $(\tilde w,\lambda)$ and a~function $\lambda(t)$ of $t$ such that
\begin{gather*}
\mathcal{L}_\mathrm{KP}(\tilde w;t) = f_\mathrm{KP}(\tilde w;\lambda(t)) = \sum\limits_{n=0}^{\infty}
u_n(\lambda(t))\tilde w^{1-n} = \tilde w + u_2(\lambda(t)) \tilde w^{-1} + u_3(\lambda(t)) \tilde w^{-2} + \cdots.
\end{gather*}
Then the inverse function $g_\mathrm{KP}(\tilde z;\lambda)$ of $f_\mathrm{KP}(\tilde w;\lambda)$ in $\tilde w$ satisf\/ies
the chordal L\"owner equation\footnote{The chordal L\"owner equation was f\/irst introduced in~\cite{k-s-s:68} and
rediscovered independently by Gibbons and Tsarev in~\cite{gib-tsa:99} and by Schramm
in~\cite{sch:00}.}~\eqref{chordal-loewner} ($\tilde g=g_\mathrm{KP}$, $\tilde u=u_2$) with respect to $\lambda$.
There is also a~way back and we can construct a~solution of the dispersionless KP hierarchy from a~solution of the
chordal L\"owner equation.
Similar connections of the dispersionless mKP/Toda hierarchies, the Dym hierarchy, the universal Whitham hierarchy to
the various L\"owner type equations were found in~\cite{man:04, m-ma-m:02, tak-tak:06, tak-tak:08, t-t-z:06}.

The goal of the present paper is to add another example: we show that, in a~similar sense, the one-variable reduction of
the dispersionless BKP hierarchy, a~variant of the dispersionless KP hierarchy, is characterised by the quadrant
L\"owner equation~\eqref{quadrant-loewner}, which is satisf\/ied by a~family of conformal mappings from a~quadrant slit
domain to the quadrant.
It is desirable to explain fundamental reason of such relation between dispersionless integrable hierarchies and
L\"owner type equations in complex analysis.

The other direction, namely, construction of a~solution of the dispersionless BKP hierarchy from a~solution of the
quadrant L\"owner type equation, can be generalised to the $N$-variable case.
The celebrated Gibbons--Tsarev system (see, for example,~\cite{gib-tsa:96, gib-tsa:99, ode-sok:09, ode-sok:10}) arises as
the compati\-bi\-li\-ty condition of the parameter functions in the quadrant L\"owner equation.
Our Gibbons--Tsarev system is a~particular case in~\cite{pav:06} and analysed there in more general context.

Let us comment several dif\/ferences from previously known results, which are due to our restricting ourselves to the
dispersionless BKP hierarchy specif\/ically and not studying general hydrodynamic type integrable systems comprehensively.
(Some of the comments also apply to the reduction of similar systems, which we studied earlier in~\cite{tak-tak:06, tak-tak:08, t-t-z:06}.)
\begin{itemize}\itemsep=-1pt
\item
We use the relation of the Grunsky coef\/f\/icients and the dispersionless tau function~\eqref{dhirota:grunsky}
systematically as in~\cite{tak-tak:06, tak-tak:08, t-t-z:06}.
We believe that this makes proofs transparent.
\item
The quadrant L\"owner equation~\eqref{quadrant-loewner} can be obtained if we put $q=g^2$, $B^0=2u$ and $p_i=V_i$ in
(3.1) of~\cite{pav:06}\footnote{The author is much grateful to one of the referees of the f\/irst version of the present
paper for pointing this out to him.}.
Instead of studying more general equations in~\cite{pav:06}, we identify our particular
equation~\eqref{quadrant-loewner} as the L\"owner type equation for the conformal mappings of quadrant domains, which
makes it possible to construct an explicit example of solutions by means of complex analysis in
Section~\ref{sec:example}.
\item
Although Tsarev's generalised hodograph method is mostly applied to the $(1{+}1)$- or $(2{+}1)$-dimensional systems (for example,
in~\cite{ode-sok:09, ode-sok:10, tsa:90}), there is no need to restrict the number of the independent variables of the
dispersionless hierarchy.
We show in this paper that the compatibility condition for Tsarev's generalised hodograph method is satisf\/ied for any
independent variables of the dispersionless BKP hierarchy {\em simultaneously}.
See Lemma~\ref{lem:hodograph} and its proof for details.
This also follows from the method by Kodama and Gibbons~\cite{kod-gib:89}.
\end{itemize}

This paper is organised as follows.
In Section~\ref{sec:dBKP} we review the results on the dispersionless BKP hierarchy.
Using its dispersionless Hirota equation, we prove that the one-variable reduction of the dispersionless BKP hierarchy
is described by the quadrant L\"owner equation in Section~\ref{sec:dBKP->quad-loewner}.
Conversely we prove in Section~\ref{sec:quad-loewner->dBKP} that a~solution of the dispersionless BKP hierarchy can be
constructed from a~function depending on $N$ parameters satisfying the quadrant L\"owner equations.
In Section~\ref{sec:quadrant-loewner} we explain that the quadrant L\"owner equation is obtained by ``folding'' the
chordal L\"owner equation by the square root, $w=\sqrt{\tilde w - 2u(\lambda)}$, where $2 u(\lambda)$ is the suitably
chosen centre of the folding.
This shows that the quadrant L\"owner equation is satisf\/ied by slit mappings between quadrant domains.
The last Section~\ref{sec:example} is devoted to the simplest non-trivial example.

\section{Dispersionless BKP hierarchy}
\label{sec:dBKP}

The dispersionless BKP hierarchy (the dBKP hierarchy for short) was f\/irst introduced probably by Kupershmidt
in~\cite{kup:83} as a~special case of the Kupershmidt hydrodynamic chain.
The formulation of the dBKP hierarchy as a~dispersionless limit of the BKP hierarchy by Date, Kashiwara and
Miwa~\cite{dkm:81}, which is relevant to us, was due to Takasaki~\cite{taka:93, taka:06} and subsequently studied by
Bogdanov and Konopelchenko~\cite{bog-kon:05}, Chen and Tu~\cite{che-tu:06}.
The theory can be developed almost in the same way as that of the dispersionless KP hierarchy in~\cite{tak-tak:95}.
We refer further developments from the viewpoint of the Kupershmidt hydrodynamic chain to~\cite{bla:02,kup:83} and \cite[especially \S~10]{pav:06} (and references therein).

Here we brief\/ly review necessary facts on the dBKP hierarchy, following mainly~\cite{taka:93, taka:06}.
Let $\mathcal{L}(w;t)$ be a~(formal) Laurent series in $w$ of the form
\begin{gather}
\mathcal{L}(w;t) = w + u_1(t) w^{-1} + u_2(t) w^{-3} + \cdots = \sum\limits_{n=0}^\infty u_n(t) w^{1-2n},
\qquad
u_0=1,
\label{L}
\end{gather}
the coef\/f\/icients of which depend on $t=(t_1, t_3, t_5, \dots)$.
We identify another independent va\-riab\-le~$x$ with~$t_1$.
The truncation operation $(\cdot)_{\geq 0}$ of the Laurent series is def\/ined by
\begin{gather*}
\left(\sum\limits_{n\in\mathbb{Z}} a_n w^n \right)_{\geq 0} = \sum\limits_{n\geq 0} a_n w^n.
\end{gather*}
The Poisson bracket $\{a(w;x), b(w;x)\}$ is def\/ined in the same way as the dispersionless KP hierarchy
\begin{gather*}
\{a(w;x), b(w;x)\} = \frac{\partial a}{\partial w} \frac{\partial b}{\partial x} - \frac{\partial a}{\partial x}
\frac{\partial b}{\partial w}.
\end{gather*}

The {\em dispersionless BKP hierarchy} is the system of Lax equations
\begin{gather}
\frac{\partial \mathcal{L}}{\partial t_n} = \{\mathcal{B}_n, \mathcal{L} \},
\qquad
n \ \text{is odd},
\label{lax}
\end{gather}
where $\mathcal{B}_n:= (\mathcal{L}^n)_{\geq 0}$.
We call $\mathcal{L}$ the {\em Lax function} of the dBKP hierarchy.

What we need later is the dispersionless Hirota equation of the tau function, which is equivalent to the above Lax
representation.
In order to def\/ine the tau function and also for later use, we introduce several terms in complex analysis.
(See~\cite{dur:83, pom:75} for details in the context of complex analysis.)

The {\em Faber polynomials} $\Phi_n(z)$ and the {\em Grunsky coefficients} $b_{mn}$ are def\/ined for a~function (or
a~Laurent series\footnote{Throughout this article, we use the word ``function'' for a~formal series.}) of the form
\begin{gather*}
g(z) = z + b_0 + b_1 z^{-1} + b_2 z^{-2} + \cdots
\end{gather*}
by the generating functions
\begin{gather}
\log\frac{g(z)-w}{z} = - \sum\limits_{n=1}^\infty \frac{z^{-n}}{n} \Phi_n(w),
\label{def:faber}\\
\nonumber
\log\frac{g(z)-g(w)}{z-w} = - \sum\limits_{m,n=1}^\infty b_{mn} z^{-m} w^{-n}.
 %\label{def:grunsky}
\end{gather}
The Faber polynomial $\Phi_n(w)$ is a~monic $n$-th order polynomial in~$w$, coef\/f\/icients of which are polynomials in
$b_k$'s.
The Grunsky coef\/f\/icient~$b_{mn}$ is a~polynomial in~$b_k$'s and symmetric in indices~$m$ and~$n$.
They are connected by the relation
\begin{gather}
\Phi_n(g(z)) = z^n + n \sum\limits_{m=0}^\infty b_{nm} z^{-m}.
\label{faber-grunsky}
\end{gather}
When $g(z)$ is an odd function, i.e., $b_{\text{even}}=0$, we have
\begin{gather*}
- \sum\limits_{n=1}^\infty \frac{(-z)^{-n}}{n} \Phi_n(w) = - \sum\limits_{n=1}^\infty \frac{z^{-n}}{n} \Phi_n(-w)
\end{gather*}
by setting $z\mapsto -z$ or $w\mapsto -w$ in~\eqref{def:faber}.
Hence the Faber polynomial $\Phi_n(w)$ is an odd (resp.\ even) polynomial in $w$ for odd (resp.\ even) $n$, and
consequently $\Phi_n(g(z))$ is an odd (resp.\ even) function in $z$.
Therefore the relation~\eqref{faber-grunsky} shows that $b_{nm}=0$ if $n$ and $m$ are of dif\/ferent parity.

We can pick up the Faber polynomials and the Grunsky coef\/f\/icients with odd indices by the following generating
functions
\begin{gather}
\log \frac{g(z)-w}{g(z)+w} = -2 \sum\limits_{1\leq n,\; \text{odd}} \frac{z^{-n}}{n} \Phi_n(w),
\label{faber:odd}
\\
\log\frac{g(z)-g(w)}{z-w} - \log\frac{g(z)+g(w)}{z+w} = -2 \sum\limits_{1\leq m, n,\; \text{odd}} b_{mn} z^{-m} w^{-n}.
\label{grunsky:odd}
\end{gather}
(The proof is straightforward computation.)

In these terms the dispersionless BKP hierarchy is stated in the following way.

\begin{proposition}
\label{prop:dhirota}
Let $\mathcal{L}(w;t)$ be a~Laurent series of the form~\eqref{L} and $k(z;t)= z + v_1(t)z^{-1} + v_2(t)z^{-3} + \cdots$
be its inverse function in the first variable: $\mathcal{L}(k(z;t);t)=z$, $k(\mathcal{L}(w;t);t)=w$.

The function $\mathcal{L}(w;t)$ is a~solution of the dBKP hierarchy~\eqref{lax} if and only if there exists a~function
$F(t)$ such that the Grunsky coefficients $b_{mn}(t)$ of $k(z;t)$ are expressed as
\begin{gather}
b_{mn}(t) = - \frac{2}{mn} \frac{\partial^2}{\partial t_m \partial t_n} F(t),
\label{dhirota:grunsky}
\end{gather}
for any odd $m$ and $n$.
\end{proposition}
The function $F(t)$ is called the {\em free energy} or the logarithm of the {\em tau function} $\tau_\mathrm{dBKP}(t)$:
$F(t)=\log \tau_\mathrm{dBKP}(t)$.
The proof of the proposition is the same as the corresponding statements for the dispersionless KP hierarchy in \S~3.3
of~\cite{teo:03}.

Since the f\/irst Faber polynomial $\Phi_1(w;t)$ of $k(z;t)$ is $w$, the relation~\eqref{faber-grunsky} means
\begin{gather}
k(z;t) = z + \sum\limits_{1\leq m,\; \text{odd}} b_{1m}(t) z^{-m}.
\label{k=z+bz}
\end{gather}
Substituting this expression into~\eqref{grunsky:odd}, the above condition~\eqref{dhirota:grunsky} is equivalent to the
following equation
\begin{gather}
4 D(z_1) D(z_2) F(t)
= \log\left( \frac{(z_1 - 2\partial_{t_1} D(z_1) F(t)) - (z_2 - 2\partial_{t_1} D(z_2) F(t))} {z_1 - z_2} \right)
\nonumber
\\
\phantom{4 D(z_1) D(z_2) F(t)=}
{}- \log\left( \frac{(z_1 - 2\partial_{t_1} D(z_1) F(t)) + (z_2 - 2\partial_{t_1} D(z_2) F(t))} {z_1 + z_2} \right),
\label{dhirota:diff-eq}
\end{gather}
where the operator $D(z)$ is def\/ined by
\begin{gather*}
D(z) = \sum\limits_{1\leq n,\; \text{odd}} \frac{z^{-n}}{n} \frac{\partial}{\partial t_n}.
\end{gather*}
Equation~\eqref{dhirota:diff-eq} is the {\em dispersionless Hirota equation} for the dBKP hierarchy f\/irst obtained
in~\cite{bog-kon:05}
(see also~\cite{taka:06}).

\section{From the dBKP hierarchy to the quadrant L\"owner equation}
\label{sec:dBKP->quad-loewner}

One of the main theorems of the present paper is the following.

\begin{theorem}
\label{thm:dBKP->loewner}
Suppose that $\mathcal{L}(w;t)$ is a~solution of the dBKP hierarchy whose dependence on $t=(t_1, t_3,\dots)$ is only
through a~single variable $\lambda$.
Namely, there exists a~function $f(w;\lambda)$ of $w$ and $\lambda$ of the form
\begin{gather*}
f(w;\lambda)= \sum\limits_{n=0}^{\infty} u_n(\lambda)w^{1-2n},
\qquad
u_0=1
%\label{f}
\end{gather*}
and a~function $\lambda(t)$ of $t$ such that
\begin{gather*}
\mathcal{L}(w;t) = f(w;\lambda(t)).
%\label{1-var-L}
\end{gather*}
We assume $du_1/d\lambda \neq 0$ and $\partial\lambda/\partial t_1 \neq 0$.
Let a~series $g(z;\lambda)$ of the form
\begin{gather}
g(z;\lambda) = \sum\limits_{n=0}^{\infty} v_n(\lambda)z^{1-2n},
\qquad
 v_0=1, \qquad v_1=-u_1
\label{g}
\end{gather}
be the inverse function of $f(w;\lambda)$ in the first variable: $g(f(w;\lambda);\lambda)=w$,
$f(g(z;\lambda);\lambda)=z$.
Then $g(z;\lambda)$ satisfies the following equation with respect to $\lambda$
\begin{gather}
\frac{\partial g}{\partial \lambda} = \frac{g}{V^2 - g^2} \frac{du}{d\lambda}.
\label{quadrant-loewner}
\end{gather}
Here $V=V(\lambda)$ is a~function of $\lambda$, not depending on $z$, and $u(\lambda)=u_1(\lambda)=-v_1(\lambda)$.

The function $\lambda(t)$ is characterised by the following system
\begin{gather}
\frac{\partial\lambda}{\partial t_n} = \chi_n(\lambda) \frac{\partial\lambda}{\partial t_1}
\qquad
\text{for odd $n$.}
\label{dlambda/dt}
\end{gather}
The coefficient $\chi_n(\lambda)$ is defined by $\chi_n(\lambda):=\Phi'_n(V(\lambda);\lambda)$, where
$\Phi_n(w;\lambda)$ is the $n$-th Faber polynomial of $g(z;\lambda)$ and $\Phi'_n=\partial \Phi_n/\partial w$.
\end{theorem}

We call the equation~\eqref{quadrant-loewner} the {\em quadrant L\"owner equation} by the reason which we shall explain
in Section~\ref{sec:quadrant-loewner}.
As in the case of the chordal L\"owner equation, we call the function $V(\lambda)$ the {\em driving function}.

\begin{remark}
As is mentioned in Section~\ref{sec:intro}, the quadrant L\"owner equation~\eqref{quadrant-loewner} can be obtained if
we put $q=g^2$, $B^0=2u$ and $p_i=V_i$ in (3.1) of~\cite{pav:06}.
Correspondingly, the system~\eqref{dlambda/dt} for $n=3$ is (2.20) of~\cite{pav:06} ($\beta=2$, $t_1\mapsto x$, $t_3
\mapsto t$).
The author thanks the referee for this reference.
\end{remark}

\begin{proof}
We follow the proof of Proposition 5.1 of~\cite{t-t-z:06}.
In this situation $g(z;t)$ is $k(z;t)$ in~\eqref{k=z+bz} and expressed by the free energy as
\begin{gather*}
g(z;t) = z - 2 \partial_{t_1} D(z) F.
%\label{g=z-ddF}
\end{gather*}
Therefore $D(z_1) g(z_2;t) = D(z_2) g(z_1;t)$, from which follows $-D(z) u(\lambda(t)) = \partial_{t_1} g(z;\lambda(t))
= \partial_\lambda g(z;\lambda) \partial_{t_1}\lambda(t)$.
Thus
\begin{gather}
D(z)\lambda(t) = -\partial_\lambda g(z;\lambda) \frac{\partial_{t_1}\lambda(t)}{\partial_\lambda u(\lambda(t))}.
\label{D(z)lambda}
\end{gather}
Since $2D(z)\partial_{t_1}F(t)=z-g(z;\lambda(t))$, the above equation and the dispersionless Hirota
equation~\eqref{dhirota:diff-eq} dif\/ferentiated by $t_1$ imply
\begin{gather*}
2 \partial_\lambda g(z_1;\lambda) \partial_\lambda g(z_2;\lambda)
= \left( \frac{\partial_\lambda g(z_1;\lambda) - \partial_\lambda g(z_2;\lambda)} {g(z_1;\lambda) - g(z_2;\lambda)} -
\frac{\partial_\lambda g(z_1;\lambda) + \partial_\lambda g(z_2;\lambda)} {g(z_1;\lambda) + g(z_2;\lambda)} \right)
\frac{d u}{d \lambda}.
\end{gather*}
This can be rewritten as
\begin{gather*}
g(z_1;\lambda)^2 + \frac{g(z_1;\lambda)}{\partial_\lambda g(z_1;\lambda)} \frac{du}{d\lambda} = g(z_2;\lambda)^2 +
\frac{g(z_2;\lambda)}{\partial_\lambda g(z_2;\lambda)} \frac{du}{d\lambda},
%\label{dbkp-loewner:temp}
\end{gather*}
which means that both-hand sides do not depend neither on $z_1$ nor on $z_2$.
Def\/ining a~func\-tion~$V(\lambda)$ of~$\lambda$ by
\begin{gather*}
V(\lambda)^2:= g(z;\lambda)^2 + \frac{g(z;\lambda)}{\partial_\lambda g(z;\lambda)} \frac{du}{d\lambda},
%\label{def:V}
\end{gather*}
we obtain the equation~\eqref{quadrant-loewner}.
Substituting it into~\eqref{D(z)lambda}, we have
\begin{gather}
\sum\limits_{1\leq n,\; \text{odd}} \frac{z^{-n}}{n} \frac{\partial \lambda}{\partial t_n} =
\frac{g(z;\lambda(t))}{g(z;\lambda(t))^2 - V(\lambda(t))^2} \frac{\partial \lambda}{\partial t_1}.
\label{dlambda/dt:temp}
\end{gather}
Since the derivative of~\eqref{faber:odd} with respect to $w$ gives
\begin{gather}
\frac{g(z)}{g(z)^2 - w^2} = \sum\limits_{1\leq n,\;\text{odd}} \frac{z^{-n}}{n} \Phi'_n(w),
\label{faber':odd}
\end{gather}
the right-hand side of~\eqref{dlambda/dt:temp} is equal to
\begin{gather*}
\sum\limits_{1\leq n,\;\text{odd}}
\frac{z^{-n}}{n} \Phi'_n\big(V(\lambda(t));\lambda(t)\big) \frac{\partial \lambda}{\partial t_1}.
\end{gather*}
Hence equation~\eqref{dlambda/dt} follows from~\eqref{dlambda/dt:temp}.
\end{proof}

\section{From the quadrant L\"owner equation to the dBKP hierarchy}
\label{sec:quad-loewner->dBKP}

In this section we prove the converse statement of Theorem~\ref{thm:dBKP->loewner} in a~generalised form ($N$-variable
diagonal reduction).

Let $f(w;{\boldsymbol \lambda})$ be a~function of the following form
\begin{gather*}
f(w;{\boldsymbol \lambda})= \sum\limits_{n=0}^{\infty} u_n({\boldsymbol \lambda})w^{1-2n},
\qquad
u_0=1,
%\label{f:N}
\end{gather*}
where ${\boldsymbol \lambda}=(\lambda_1,\dots,\lambda_N)$ is a~set of $N$ variables.
We denote its inverse function in~$w$ by $g(z;{\boldsymbol \lambda})$: $f(g(z;{\boldsymbol \lambda});{\boldsymbol
\lambda})=z$, $g(f(w;{\boldsymbol \lambda});{\boldsymbol \lambda}) = w$.

Let us consider a~system of dif\/ferential equations for $g(z;{\boldsymbol \lambda})$ of the type~\eqref{quadrant-loewner}
in each va\-riab\-le~$\lambda_i$, namely, the quadrant L\"owner equations
\begin{gather}
\frac{\partial g}{\partial \lambda_i}(z;{\boldsymbol \lambda}) = \frac{g(z;{\boldsymbol \lambda})}{V_i^2({\boldsymbol
\lambda}) - g^2(z;{\boldsymbol \lambda})} \frac{\partial u}{\partial \lambda_i}({\boldsymbol \lambda}),
\label{quadrant-loewner:N}
\end{gather}
with a~driving function $V_i$ for $i=1,\dots,N$, where $u=u_1$.
Or equivalently we consider a~system of linear partial dif\/ferential equations for~$f$
\begin{gather}
\frac{\partial f}{\partial \lambda_i}(w;{\boldsymbol \lambda}) = A_i f(w;{\boldsymbol \lambda}),
\qquad
A_i:= - \frac{w}{V_i^2({\boldsymbol \lambda}) - w^2} \frac{\partial u}{\partial \lambda_i}({\boldsymbol \lambda})
\frac{\partial}{\partial w}.
\label{quadrant-loewner:N:f}
\end{gather}
When $N\geqq 2$, the compatibility conditions  $ [\partial_{\lambda_i} - A_i, \partial_{\lambda_j} - A_j]=0 $  of this
linear system boil down to the following equations for $V_i$ (or $V_i^2$) and $u$
\begin{gather}
\frac{\partial V_i^2}{\partial \lambda_j}=\frac{2 V_i^2}{V_j^2 - V_i^2} \frac{\partial u}{\partial \lambda_j},
\label{dV/dl}
\\
\frac{\partial^2 u}{\partial \lambda_i \, \partial \lambda_j}=\frac{2(V_i^2 + V_j^2)}{(V_i^2 - V_j^2)^2}
\frac{\partial u}{\partial \lambda_i} \frac{\partial u}{\partial \lambda_j}
\label{du/dldl}
\end{gather}
for $i,j=1,\dots,N$, $i\neq j$.
This is the {\em Gibbons--Tsarev system} in our case, which is well-known in the literature.
(See the remark below.) We assume that these conditions hold and that $g$ and $f$ satisfy~\eqref{quadrant-loewner:N}
and~\eqref{quadrant-loewner:N:f} respectively.

\begin{remark}
This Gibbons--Tsarev system is a~specialisation of (3.2) of~\cite{pav:06}.
For general reference of the Gibbons--Tsarev systems, see~\cite{gib-tsa:96, gib-tsa:99, ode-sok:09, ode-sok:10}.
The author thanks the referee for these references.
\end{remark}

The converse statement to Theorem~\ref{thm:dBKP->loewner} is the following.

\begin{theorem}
\label{thm:loewner->dBKP}
Suppose each $\lambda_i(t)$ satisfies the following analogue of~\eqref{dlambda/dt} for all odd $n$
\begin{gather}
\frac{\partial\lambda_i}{\partial t_n}(t) = \chi_{i,n}({\boldsymbol \lambda}(t)) \frac{\partial\lambda_i}{\partial
t_1}(t),
\label{dlambda/dt:N}
\end{gather}
where we define the function $\chi_{i,n}({\boldsymbol \lambda})$
by $\chi_{i,n}({\boldsymbol\lambda}):=\Phi'_n(V_i({\boldsymbol \lambda});{\boldsymbol \lambda})$, using
the derivative $\Phi'_n = \partial\Phi_n/\partial w$ of the $n$-th Faber polynomial $\Phi_n(w;{\boldsymbol \lambda})$
of $g(z;{\boldsymbol \lambda})$.

Then $\mathcal{L}(w;t):=f(w;{\boldsymbol \lambda}(t))$ is a~solution of the dBKP hierarchy.
\end{theorem}

This is the so-called {\em diagonal $N$-variable reduction}.
We shall discuss the solvability of~\eqref{dlambda/dt:N} later.

\begin{remark}%\label{rem:general-N-var-reduction}
We can consider more general $N$-variable reduction, changing the equation~\eqref{quadrant-loewner:N} to
\begin{gather*}
\frac{\partial g}{\partial \lambda_i}(z;{\boldsymbol \lambda}) = R_i(g(z;{\boldsymbol \lambda}); {\boldsymbol \lambda})
%\label{quadrant-loewner:N-general}
\end{gather*}
and correspondingly the equation~\eqref{quadrant-loewner:N:f} to
\begin{gather}
\frac{\partial f}{\partial \lambda_i}(w;{\boldsymbol \lambda}) = - R_i(w; {\boldsymbol \lambda}) \frac{\partial
f}{\partial w}(w;{\boldsymbol \lambda}),
\label{quadrant-loewner:N-general:f}
\end{gather}
where the coef\/f\/icient $R_i(w;{\boldsymbol \lambda})$ is an odd rational function with $2N$ simple poles at $w=\pm
V_j({\boldsymbol \lambda})$ ($j=1,\dots,N$) which vanishes at $\infty$: $R_i(w;{\boldsymbol \lambda})\to0$
($w\to\infty$).
In other words,
\begin{gather*}
R_i(w;{\boldsymbol \lambda}) = \sum\limits_{j=1}^N \frac{\rho_{ij}({\boldsymbol \lambda})\, w}{V_j^2({\boldsymbol
\lambda}) - w^2}.
\end{gather*}
Because of the equation~\eqref{quadrant-loewner:N-general:f}, $V_i({\boldsymbol \lambda})$ is a~critical point of $f$
under the genericity assumption that~$f$ is holomorphic around $w=V_i$.
As in~\cite[\S~3.1]{m-ma-m:02} (or in~\cite[\S~6]{tak-tak:06}), if we change the coordinates
$(\lambda_1,\dots,\lambda_N)$ to the critical values $\zeta_i:=f(V_i({\boldsymbol \lambda});{\boldsymbol \lambda})$ of
$f$ ($i=1,\dots,N$), then the system~\eqref{quadrant-loewner:N-general:f} reduces to the diagonal reduction
case~\eqref{quadrant-loewner:N:f} with respect to $(\zeta_1,\dots,\zeta_N)$.
In this context $\zeta_i$'s are the Riemann invariants.
\end{remark}

\begin{proof}[Proof of Theorem~\ref{thm:loewner->dBKP}]  This is proved in the same way as Proposition~5.5 of~\cite{t-t-z:06}.
From~\eqref{grunsky:odd} it follows that
\begin{gather}
-2 \sum\limits_{1\leq m, n,\;\text{odd}} \frac{\partial b_{mn}}{\partial t_k} z_1^{-m} z_2^{-n}
\nonumber
\\
\qquad
= \sum\limits_{i=1}^N \frac{ 2 g(z_1;{\boldsymbol \lambda}(t)) g(z_2;{\boldsymbol \lambda}(t))} {( V({\boldsymbol
\lambda}(t))^2 - g(z_1;{\boldsymbol \lambda}(t))^2 ) ( V({\boldsymbol \lambda}(t))^2 - g(z_2;{\boldsymbol \lambda}(t))^2
)} \frac{\partial u}{\partial \lambda_i} \frac{\partial\lambda_i}{\partial t_k}
\label{d(grunsky:odd)+loewner}
\end{gather}
under the assumption~\eqref{quadrant-loewner:N}.
The right-hand side is rewritten as
\begin{gather*}
2 \sum\limits_{i=1}^N \sum\limits_{1\leq m,n,\;\text{odd}} \frac{z_1^{-m}}{m} \frac{z_2^{-n}}{n} \chi_{i,m}(\lambda(t))
\chi_{i,n}(\lambda(t)) \chi_{i,k}(\lambda(t)) \frac{\partial u}{\partial \lambda_i} \frac{\partial\lambda_i}{\partial t_1}
\end{gather*}
by virtue of~\eqref{faber':odd} and~\eqref{dlambda/dt:N}.
Therefore~\eqref{d(grunsky:odd)+loewner} implies that $mn \, \partial b_{mn}/\partial t_k$ is symmetric in $(m,n,k)$,
from which follows the existence of a~function $F(t)$ satisfying~\eqref{dhirota:grunsky}.
According to Proposition~\ref{prop:dhirota}, $\mathcal{L}(w;t)=f(w;{\boldsymbol \lambda}(t))$ is a~solution of the dBKP
hierarchy.
\end{proof}

The system of the f\/irst order partial dif\/ferential equations~\eqref{dlambda/dt:N} can be solved by Tsarev's generalised
hodograph method~\cite{tsa:90}.

\begin{lemma}
\label{lem:hodograph}
Consider the following system for $R_i=R_i({\boldsymbol \lambda})$, $i=1,\dots,N$:
\begin{gather}
\frac{\partial R_i}{\partial \lambda_j} = \Gamma_{ij} (R_j-R_i),
\qquad
i,j=1,\dots,N,
\quad
i\neq j,
\label{dR/dl}
\end{gather}
where $\Gamma_{ij}$ is defined by
\begin{gather*}
\Gamma_{ij}:= \frac{V_i^2 + V_j^2}{(V_i^2 - V_j^2)^2} \frac{\partial u}{\partial \lambda_i}.
%\label{def:Gamma}
\end{gather*}
$($When $N=1$, the condition~\eqref{dR/dl} is void.$)$

$(i)$  The system~\eqref{dR/dl} is compatible in the sense of~{\rm \cite[\S~3]{tsa:90}}.

$(ii)$  Assume that $R_i({\boldsymbol \lambda})$ satisfy the system~\eqref{dR/dl}.
If $\lambda(t)$ is defined implicitly by the hodograph relation
\begin{gather}
t_1 + \sum\limits_{1\leq n,\; \text{odd}} \chi_{i,n}({\boldsymbol \lambda})t_n = R_i({\boldsymbol \lambda}),
\label{hodograph}
\end{gather}
then ${\boldsymbol \lambda}(t)$ satisfies~\eqref{dlambda/dt:N}.
\end{lemma}

This is essentially Theorem 10 of~\cite{tsa:90}, which is well-known to experts, but since the number of the independent
variables are inf\/inite in our case, we brief\/ly review the proof and check that it really works.
To handle the inf\/inite number of equations simultaneously, we make use of the generating function of the Faber
polynomials.
We also note that this lemma also follows directly from the generalised hodograph method in the slightly improved
version made by Kodama and Gibbons~\cite{kod-gib:89} for f\/initely many independent variables.

\begin{proof}
This proof is parallel to the proof of Lemma 4.1 of~\cite{tak-tak:06}.
First we prove that $\chi_{i,n}$ ($n$ is odd) satisf\/ies the same equation~\eqref{dR/dl} as $R_i$.
Since
\begin{gather*}
\frac{\partial \chi_{i,n}}{\partial \lambda_j}({\boldsymbol \lambda}) = \frac{\partial V_i}{\partial
\lambda_j}({\boldsymbol \lambda}) \frac{\partial^2 \Phi_n}{\partial w^2}(V_i({\boldsymbol \lambda});{\boldsymbol
\lambda}) + \frac{\partial^2 \Phi_n}{\partial w\, \partial \lambda_j} (V_i({\boldsymbol \lambda});{\boldsymbol
\lambda}),
\end{gather*}
the generating function for $\partial \chi_{i,n}/\partial \lambda_j$ is obtained by dif\/ferentiating~\eqref{faber':odd}.
The result is
\begin{gather*}
\sum\limits_{1\leq n,\;\text{odd}} \frac{z^{-n}}{n} \frac{\partial \chi_{i,n}}{\partial \lambda_j} = \frac{1}{(g^2 -
V_i^2)^2} \left( \frac{\partial V_i^2}{\partial\lambda_j} g - (g^2 + V_i^2) \frac{\partial g}{\partial \lambda_j}
\right).
\end{gather*}
Substituting the compatibility condition~\eqref{dV/dl} and the quadrant L\"owner equations~\eqref{quadrant-loewner:N},
we obtain
\begin{gather}
\sum\limits_{1 \leq n,\;\text{odd}} \frac{z^{-n}}{n} \frac{\partial \chi_{i,n}}{\partial \lambda_j} =
\frac{g}{(g^2-V_j^2)(g^2-V_i^2)} \frac{V_i^2 + V_j^2}{V_j^2 - V_i^2} \frac{\partial u}{\partial \lambda_j}.
\label{gen-func:dchi/dl}
\end{gather}
On the other hand, it follows from~\eqref{faber':odd} and the def\/inition of $\chi_{i,n}$ that
\begin{gather}
\sum\limits_{1\leq n,\;\text{odd}} \frac{z^{-n}}{n} (\chi_{j,n}-\chi_{i,n})
= \frac{g\big(V_j^2-V_i^2\big)}{\big(g^2-V_j^2\big)\big(g^2-V_i^2\big)}.
\label{gen-func:chi-chi}
\end{gather}
Comparing~\eqref{gen-func:dchi/dl} and~\eqref{gen-func:chi-chi}, we have
\begin{gather}
\frac{\partial \chi_{i,n}}{\partial \lambda_j} = \Gamma_{ij} (\chi_{j,n} - \chi_{i,n}),
\qquad
\text{namely,}
\qquad
\Gamma_{ij} = \frac{\partial_{\lambda_j} \chi_{i,n}}{\chi_{j,n} - \chi_{i,n}},
\label{dchi/dl=Gamma(chi-chi)}
\end{gather}
for odd $n$ and $i,j=1,\dots,N$, $i\neq j$.
An important point is that this equation holds {\em for all $n$ simultaneously}.
Namely, the coef\/f\/icient of the equation~\eqref{dchi/dl=Gamma(chi-chi)} does not depend on $n$.

$(i)$  Lengthy computation with the help of~\eqref{dV/dl} and~\eqref{du/dldl} shows that
\begin{gather*}
\frac{\partial \Gamma_{ij}}{\partial \lambda_k} = \frac{2 (V_i^2 (V_j^2 + V_k^2) + 2 V_j^2 V_k^2)} {(V_j^2 -
V_i^2)(V_k^2 - V_i^2)(V_k^2 - V_j^2)^2} \frac{\partial u}{\partial \lambda_j} \frac{\partial u}{\partial \lambda_k}.
\end{gather*}
The right-hand side is symmetric in $(j,k)$.
Hence we have
\begin{gather*}
\frac{\partial \Gamma_{ij}}{\partial \lambda_k} = \frac{\partial \Gamma_{ik}}{\partial \lambda_j}.
\end{gather*}
This together with the expression~\eqref{dchi/dl=Gamma(chi-chi)} is the compatibility condition of~\eqref{dR/dl}, as
shown in \S~3 of~\cite{tsa:90}.

$(ii)$  By dif\/ferentiating the relation~\eqref{hodograph} by $t_1$ and $t_k$ ($k$ is odd) we obtain
\begin{gather}
\sum\limits_{j=1}^N M_{ij} \frac{\partial \lambda_j}{\partial t_1} = 1,
\qquad
\sum\limits_{j=1}^N M_{ij} \frac{\partial \lambda_j}{\partial t_k} = \chi_{i,k},
\label{d(hodograph)}
\end{gather}
where
\begin{gather*}
M_{ij}:= \frac{\partial R_i}{\partial \lambda_j} - \sum\limits_{1\leq n,\; \text{odd}} \frac{\partial \chi_{i,n}}{\partial
\lambda_j} t_n.
%\label{def:Mij}
\end{gather*}
Because of~\eqref{dR/dl},~\eqref{dchi/dl=Gamma(chi-chi)} and the hodograph relation~\eqref{hodograph} the above
expression becomes
\begin{gather*}
M_{ij}=\Gamma_{ij}(R_j-R_i) - \sum\limits_{1\leq n,\; \text{odd}} \Gamma_{ij} (\chi_{j,n}-\chi_{i,n}) t_n
\\
\phantom{M_{ij}}= \Gamma_{ij}\bigl( (R_j - \sum\limits_{1\leq n,\; \text{odd}} \chi_{j,n} t_n) - (R_i - \sum\limits_{1\leq n,\; \text{odd}}
\chi_{i,n} t_n) \bigr) = 0,
\end{gather*}
if $i\neq j$.
(The fact that the coef\/f\/icient $\Gamma_{ij}$ does not depend on $n$ is essential here.) Therefore~\eqref{d(hodograph)}
reduces to
\begin{gather*}
M_{ii} \frac{\partial \lambda_i}{\partial t_1} = 1,
\qquad
M_{ii} \frac{\partial \lambda_i}{\partial t_k} = \chi_{i,k}.
\end{gather*}
This proves~\eqref{dlambda/dt:N}.
\end{proof}

\section{Quadrant L\"owner equation}
\label{sec:quadrant-loewner}

In this section we show that the quadrant L\"owner equation~\eqref{quadrant-loewner}
\begin{gather*}
\frac{\partial g}{\partial \lambda}(z;\lambda) = \frac{g(z;\lambda)}{V(\lambda)^2 - g(z;\lambda)^2}
\frac{du}{d\lambda}(\lambda)
%\label{quadrant-loewner:}
\end{gather*}
is satisf\/ied by slit mappings between quadrant domains.

In fact this equation is connected to the chordal L\"owner equation as follows.
\begin{proposition}
\label{prop:quad-chordal}
$(i)$  If the function $g(z;\lambda)$ of the form~\eqref{g} satisfies the equation~\eqref{quadrant-loewner} $(u=-v_1)$,
then $\tilde g(\tilde z;\lambda):= g(\sqrt{\tilde z};\lambda)^2 - 2v_1(\lambda)$ satisfies the hydrodynamic
normalisation condition
\begin{gather}
\tilde g(\tilde z; \lambda) = \tilde z + \tilde u(\lambda) \tilde z^{-1} + O\big(\tilde z^{-2}\big)
\label{gtilde:normalised}
\end{gather}
and the chordal L\"owner equation
\begin{gather}
\frac{\partial \tilde g}{\partial \lambda} = \frac{1}{\tilde g - U} \frac{d\tilde u}{d\lambda}.
\label{chordal-loewner}
\end{gather}
Here the driving function $U=U(\lambda)$ is
%\begin{gather*}
$U(\lambda) = V(\lambda)^2-2v_1(\lambda)$.
%\label{U=V2-2v1}
%\end{gather*}
Moreover, $\tilde u(\lambda)$ satisfies
\begin{gather}
\frac{d\tilde u}{d\lambda} = - 2V^2 \frac{du}{d\lambda}.
\label{dtildeu=-2V2du}
\end{gather}

$(ii)$  Conversely, let $\tilde g(\tilde z;\lambda) = \tilde z + \tilde u(\lambda) \tilde z^{-1} + \cdots$ be a~solution
of the chordal L\"owner equation~\eqref{chordal-loewner} with the driving function $U(\lambda)$.
Let $V(\lambda)$ be a~solution of the following ordinary differential equation
\begin{gather}
\frac{d}{d\lambda} V^4 - 2 \frac{dU}{d\lambda} V^2 = 2 \frac{d\tilde u}{d\lambda}.
\label{V:diff-eq}
\end{gather}
If we define $u(\lambda)$ by
\begin{gather}
u(\lambda) = \frac{1}{2}\big(U(\lambda) - V(\lambda)^2\big),
\label{u=(U-V2)/2}
\end{gather}
then $g(z;\lambda):= \sqrt{\tilde g(z^2;\lambda) - 2u(\lambda)}$ is of the form~\eqref{g} and satisfies
equation~\eqref{quadrant-loewner}.
$($Exactly speaking, we can choose a~branch of the square root so that $g(z;\lambda)$ is of the form~\eqref{g} and
satisfies~\eqref{quadrant-loewner}.$)$
\end{proposition}

\begin{figure}[t] \centering
\includegraphics{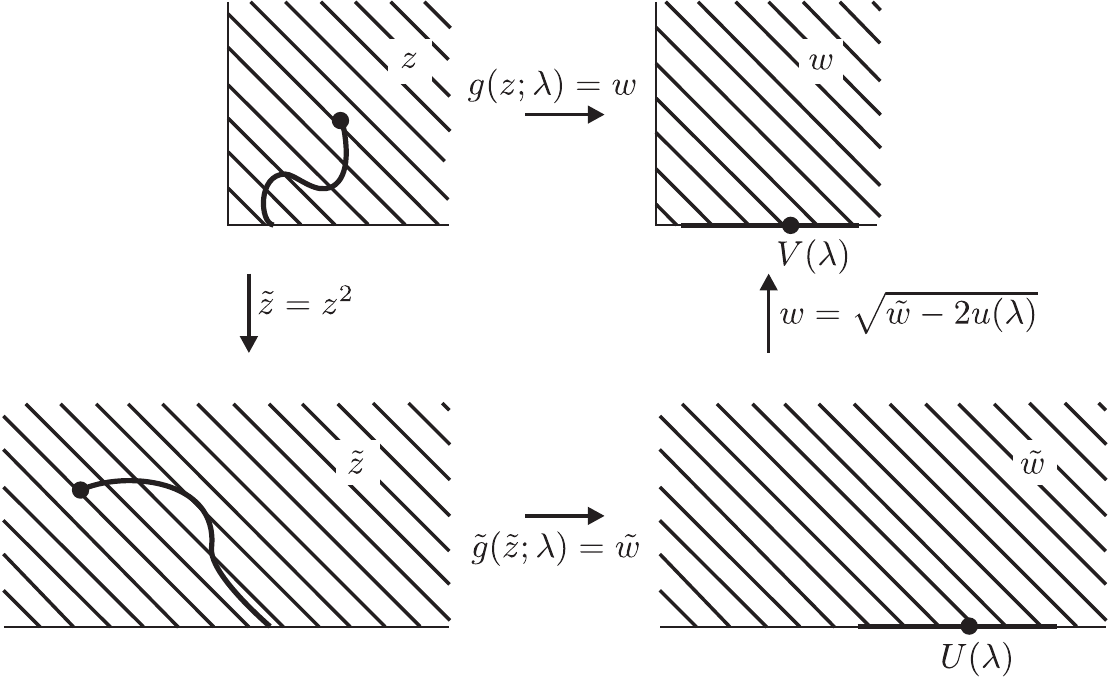}

\caption{Conformal mapping from a~slit domain to the quadrant.}\label{fig:quadrant-loewner}
%{\small $U(\lambda)$ \quad $\tilde z$\quad $\tilde w$ \quad $\tilde g(\tilde z;\lambda)=\tilde w$
%\quad $\tilde z=z^2$ \quad $w=\sqrt{\tilde w -2u(\lambda)}$ \quad $V(\lambda)$ \quad $g(z;\lambda)= w$ \quad $z$\quad $w$}
\end{figure}

As is well known (cf., for example,~\cite{a-b-c-d:10}), a~one-parameter family of conformal mappings from a~slit domain
to the upper half plane satisf\/ies the chordal L\"owner equation\footnote{Recently a~proof without advanced techniques
was given in~\cite{d-mo-gum:13}.}.
Hence the equation~\eqref{quadrant-loewner} is essentially an equation for conformal mappings from a~slit domain to the
quadrant (Fig.~\ref{fig:quadrant-loewner}).
By this reason we call~\eqref{quadrant-loewner} the {\em quadrant L\"owner equation}.

Actually, we can ``fold'' the function $\tilde g(z^2;\lambda)$ at any point $c(\lambda)$ on the real axis instead of
$2u(\lambda)$ to obtain the conformal mapping to the quadrant: $g_c(z;\lambda)=\sqrt{\tilde g(z^2;\lambda) -
c(\lambda)}$.
As is shown in the proof, the shift by $2u(\lambda)$ is necessary to normalise $\tilde g$ by the hydrodynamic
normalisation condition.

\begin{proof}
Both statements are consequences of straightforward computation.

For example, as for $(i)$, it is easy to check that $\tilde g(\tilde z;\lambda)$ satisf\/ies the normalisation
condition~\eqref{gtilde:normalised}, since
\begin{gather*}
\begin{split}
&\tilde g(\tilde z;\lambda)=\left( \sum\limits_{n=0}^\infty v_n(\lambda) \bigl(\sqrt{\tilde z}\bigr)^{1-2n} \right)^2
- 2 v_1(\lambda)
= \tilde z \left(\sum\limits_{n=0}^\infty v_n(\lambda) \tilde z^{-n}\right)^2 - 2 v_1(\lambda)
\\
&\phantom{\tilde g(\tilde z;\lambda)}
= \tilde z + (2v_2+v_1^2) \tilde z^{-1} + O\big(\tilde z^{-2}\big).
\end{split}
\end{gather*}
Next, the chordal L\"owner equation can be checked as follows
(recall that $v_1=-u$)
\begin{gather*}
\frac{\partial \tilde g}{\partial \lambda}(\tilde z;\lambda)=2g\big(\sqrt{\tilde z};\lambda\big) \frac{\partial g}{\partial
\lambda}\big(\sqrt{\tilde z};\lambda\big) - 2 \frac{dv_1}{d\lambda}
= \left(\frac{2 g^2}{V^2 - g^2} + 2\right) \frac{du}{d\lambda} = \frac{2V^2}{V^2-g^2} \frac{du}{d\lambda}.
\end{gather*}
Hence summarising, we have
\begin{gather}
\frac{\partial \tilde g}{\partial \lambda} = \frac{2V^2}{U-\tilde g} \frac{du}{d\lambda}.
\label{chordal-loewner:temp}
\end{gather}
By comparing the coef\/f\/icients of $\tilde z^{-1}$ we have~\eqref{dtildeu=-2V2du}, which implies~\eqref{chordal-loewner}
from~\eqref{chordal-loewner:temp}.

The proof of $(ii)$ is similar computation of a~Laurent series and derivation.
The point is that the conditions~\eqref{V:diff-eq} and~\eqref{u=(U-V2)/2} lead to the equation~\eqref{dtildeu=-2V2du}.
\end{proof}

\section{Example}
\label{sec:example}

Here, as an application of Theorem~\ref{thm:loewner->dBKP} ($N=1$) and Proposition~\ref{prop:quad-chordal}, we construct
an example of a~solution of the dBKP hierarchy, starting from a~solution of the chordal L\"owner equation.

The following function $\tilde g(\tilde z;\lambda)$ is a~solution of the chordal L\"owner equation with the driving
function $U(\lambda)=U$ (a constant)
\begin{gather}
\tilde g(\tilde z;\lambda) = U + \sqrt{(\tilde z-U)^2+2\lambda}
\nonumber
\\
\phantom{\tilde g(\tilde z;\lambda)}
= U + \tilde z \sqrt{ 1 - 2U\tilde z^{-1} + (U^2 + 2\lambda)\tilde z^{-2}} = \tilde z + \lambda \tilde z^{-1} + \lambda
U \tilde z^{-2} + \cdots.
\label{tildeg}
\end{gather}
(The branch of the square root is chosen so that it has the expansion of the above form.) See, for example, Appendix
A.1.1 of~\cite{t-t-z:06}.
This maps the slit domain  $ \mathbb{H} \backslash \{U+i\alpha \mid \alpha \in [0, \sqrt{2\lambda}]\} $  to the
upper half plane $\mathbb{H}$.
Since the function $\tilde u(\lambda)$ is equal to $\lambda$, as is seen from the expansion~\eqref{tildeg}, the
dif\/ferential equation~\eqref{V:diff-eq} becomes $\frac{d}{d\lambda} V^4 = 2$, which is readily solved:
\begin{gather}
V^2(\lambda) = (2\lambda+c)^{1/2},
\label{V:example}
\end{gather}
where $c$ is the integration constant and the branch of the square root can be chosen arbitrarily.
Hence $u(\lambda)=(U-V(\lambda)^2)/2=(U-(2\lambda+c)^{1/2})/2$ and
\begin{gather*}
g(z;\lambda)=\sqrt{\tilde g(z^2;\lambda) - 2u(\lambda)}
= \sqrt{ \sqrt{ (z^2-U)^2 + 2\lambda } + (2\lambda+c)^{1/2} }
\\
\phantom{g(z;\lambda)}
= z \sqrt{ \sqrt{ (1-Uz^{-2})^2 + 2\lambda z^{-4} } + (2\lambda+c)^{1/2} z^{-2} }
\\
\phantom{g(z;\lambda)}
= z + \frac{-U+(2\lambda+c)^{1/2}}{2} z^{-1}
+ \left( \frac{\lambda}4 - \frac{U^2}{8} + \frac{U(2\lambda+c)^{1/2}}{4}-\frac{c}{8} \right) z^{-3} + \cdots.
%\label{g:example}
\end{gather*}
Therefore the Faber polynomials $\Phi_1(w;\lambda)$ and $\Phi_3(w;\lambda)$ are explicitly calculated as
\begin{gather}
\Phi_1(w;\lambda) = w,
\qquad
\Phi_3(w;\lambda) = w^3 - 3\frac{-U+(2\lambda+c)^{1/2}}{2} w.
\label{faber:example}
\end{gather}
Using~\eqref{V:example} and~\eqref{faber:example}, we have
\begin{gather*}
\chi_1(\lambda)=1,
\qquad
\chi_3(\lambda)=\frac{3}{2} (2\lambda+c)^{1/2} + \frac{3}{2}U.
%\label{chi:example}
\end{gather*}
Applying Lemma~\ref{lem:hodograph} with $R(\lambda)=0$, $\lambda(t)_{t_5=t_7=\cdots=0}$ is determined by the relation
\begin{gather*}
t_1 + \frac{3}{2} (U+(2\lambda+c)^{1/2}) t_3 = 0,
\end{gather*}
namely,
\begin{gather*}
\lambda(t_1, t_3) = \frac{1}{2} \left( \left( \frac{2t_1}{3t_3} + U \right)^2 - c \right).
%\label{lambda:example}
\end{gather*}
The inverse function $f(w;\lambda)$ of $g(z;\lambda)$ is
\begin{gather*}
f(w;\lambda)=\sqrt{ U + \sqrt{ (w^2 - (2\lambda+c)^{1/2})^2 - 2 \lambda } }
\\
\phantom{f(w;\lambda)}= w + \frac{U - (2\lambda+c)^{1/2}}{2} w^{-1} + \left( -\frac{3\lambda}{4} - \frac{U^2}{8}
+ \frac{U(2\lambda+c)^{1/2}}{4} - \frac{c}{8} \right) w^{-3} + \cdots.
%\label{f:example}
\end{gather*}
Thus we obtain a~solution of the dBKP hierarchy (with respect to $t_1$ and $t_3$) by Theorem~\ref{thm:dBKP->loewner}:
\begin{gather*}
\mathcal{L}(w;t)|_{t_{2n+1}=0 (n\geq 2)} = \sqrt{ U + \sqrt{ \left( w^2 - \frac{2t_1}{3t_3} - U \right)^2 -
\left(\frac{2t_1}{3t_3} + U \right)^2 + c } }.
\end{gather*}

\begin{remark}
The ordinary dif\/ferential equation~\eqref{V:diff-eq} for $V^2$ is a~special case of Chini's equation (see \cite[C.I.55, p.~303]{kam:59}; $x=\lambda$, $y=V(\lambda)^2$, $n=-1$, $f(x)=d\tilde u/d\lambda$, $g(x)=0$,
$h(x)=dU/d\lambda$).
It can be solved explicitly only in special cases.
The above example is one of them.
\end{remark}

\subsection*{Acknowledgements}

The author would like to thank Michio Jimbo and Saburo Kakei for interest to this work and for hospitality during his
stay in Rikkyo University, where this work was completed.
He also express gratitude to Anton Zabrodin for interest and comments.
Special thanks of the author are to the referees of the f\/irst version of this work, who suggested to study the
$N$-variable reduction, which was absent, and informed many references.
This study was carried out within ``The National Research University Higher School of
Economics'' Academic Fund Program in 2013--2014, research grant No.~12-01-0075.

\pdfbookmark[1]{References}{ref}
\LastPageEnding


\begin{thebibliography}{99}
\footnotesize\itemsep=-1pt

\bibitem{a-b-c-d:10}
Abate M., Bracci F., Contreras M.D., D{\'\i}az-Madrigal S., The evolution of
  Loewner's dif\/ferential equations, \textit{Eur. Math. Soc. Newsl.} \textbf{78}
  (2010), 31--38.

\bibitem{bla:02}
B{\l}aszak M., Classical {$R$}-matrices on {P}oisson algebras and related
  dispersionless systems, \href{http://dx.doi.org/10.1016/S0375-9601(02)00421-8}{\textit{Phys. Lett.~A}} \textbf{297} (2002), 191--195.

\bibitem{bog-kon:05}
Bogdanov L.V., Konopelchenko B.G., On dispersionless {BKP} hierarchy and its
  reductions, \href{http://dx.doi.org/10.2991/jnmp.2005.12.s1.6}{\textit{J.~Nonlinear Math. Phys.}} \textbf{12} (2005), suppl.~1,
  64--73, \href{http://arxiv.org/abs/nlin.SI/0411046}{nlin.SI/0411046}.

\bibitem{che-tu:06}
Chen Y.-T., Tu M.-H., A note on the dispersionless {BKP} hierarchy,
  \href{http://dx.doi.org/10.1088/0305-4470/39/24/003}{\textit{J.~Phys.~A: Math. Gen.}} \textbf{39} (2006), 7641--7655.

\bibitem{dkm:81}
Date E., Kashiwara M., Miwa T., Transformation groups for soliton equations.
  {II}.~{V}ertex operators and {$\tau$} functions, \href{http://dx.doi.org/10.3792/pjaa.57.387}{\textit{Proc. Japan Acad.
  Ser.~A Math. Sci.}} \textbf{57} (1981), 387--392.

\bibitem{d-mo-gum:13}
Del~Monaco A., Gumenyuk P., Chordal Loewner equation, \href{http://arxiv.org/abs/1302.0898}{arXiv:1302.0898}.

\bibitem{dur:83}
Duren P.L., Univalent functions, \textit{Grundlehren der Mathematischen
  Wissenschaften}, Vol.~259, Springer-Verlag, New York, 1983.

\bibitem{gib-tsa:96}
Gibbons J., Tsarev S.P., Reductions of the {B}enney equations, \href{http://dx.doi.org/10.1016/0375-9601(95)00954-X}{\textit{Phys.
  Lett.~A}} \textbf{211} (1996), 19--24.

\bibitem{gib-tsa:99}
Gibbons J., Tsarev S.P., Conformal maps and reductions of the {B}enney
  equations, \href{http://dx.doi.org/10.1016/S0375-9601(99)00389-8}{\textit{Phys. Lett.~A}} \textbf{258} (1999), 263--271.

\bibitem{kam:59}
Kamke E., Dif\/ferentialgleichungen. {L}\"osungsmethoden und {L}\"osungen,
  \textit{Mathematik und ihre Anwendungen in Physik und Technik, Reihe~A}, Bd.~18,
  Akademische Verlagsgesellschaft Geest \& Portig K.-G., Leipzig, 1959.

\bibitem{kod-gib:89}
Kodama Y., Gibbons J., A method for solving the dispersionless {KP} hierarchy
  and its exact solutions.~{II}, \href{http://dx.doi.org/10.1016/0375-9601(89)90255-7}{\textit{Phys. Lett.~A}} \textbf{135} (1989),
  167--170.

\bibitem{k-s-s:68}
Kufarev P.P., Sobolev V.V., Spory{\v{s}}eva L.V., A certain method of
  investigation of extremal problems for functions that are univalent in the
  half-plane, \textit{Trudy Tomsk. Gos. Univ. Ser. Meh.-Mat.} \textbf{200}
  (1968), 142--164.

\bibitem{kup:83}
Kupershmidt B.A., Deformations of integrable systems, \textit{Proc. Roy. Irish
  Acad. Sect.~A} \textbf{83} (1983), 45--74.

\bibitem{man:04}
Ma{\~n}as M., {$S$}-functions, reductions and hodograph solutions of the
  {$r$}th dispersionless modif\/ied {KP} and {D}ym hierarchies,
  \href{http://dx.doi.org/10.1088/0305-4470/37/46/007}{\textit{J.~Phys.~A: Math. Gen.}} \textbf{37} (2004), 11191--11221,
  \href{http://arxiv.org/abs/nlin.SI/0405028}{nlin.SI/0405028}.

\bibitem{m-ma-m:02}
Ma{\~n}as M., Mart{\'{\i}}nez~Alonso L., Medina E., Reductions and hodograph
  solutions of the dispersionless {KP} hierarchy, \href{http://dx.doi.org/10.1088/0305-4470/35/2/316}{\textit{J.~Phys.~A: Math.
  Gen.}} \textbf{35} (2002), 401--417.

\bibitem{ode-sok:09}
Odesskii A.V., Sokolov V.V., Systems of Gibbons--Tsarev type and integrable
  $3$-dimensional models, \href{http://arxiv.org/abs/0906.3509}{arXiv:0906.3509}.

\bibitem{ode-sok:10}
Odesskii A.V., Sokolov V.V., Integrable (2+1)-dimensional systems of
  hydrodynamic type, \href{http://dx.doi.org/10.1007/s11232-010-0043-1}{\textit{Theoret. and Math. Phys.}} \textbf{163} (2010),
  549--586, \href{http://arxiv.org/abs/1009.2778}{arXiv:1009.2778}.

\bibitem{pav:06}
Pavlov M.V., The {K}upershmidt hydrodynamic chains and lattices, \href{http://dx.doi.org/10.1155/IMRN/2006/46987}{\textit{Int.
  Math. Res. Not.}} \textbf{2006} (2006), Art.~ID~46987, 43~pages,
  \href{http://arxiv.org/abs/nlin.SI/0604049}{nlin.SI/0604049}.

\bibitem{pom:75}
Pommerenke C., Univalent functions, \textit{Studia Mathematica/Mathematische
  Lehrb{\"u}cher}, Band~XXV, Vandenhoeck \& Ruprecht, G\"ottingen, 1975.

\bibitem{sch:00}
Schramm O., Scaling limits of loop-erased random walks and uniform spanning
  trees, \href{http://dx.doi.org/10.1007/BF02803524}{\textit{Israel~J. Math.}} \textbf{118} (2000), 221--288,
  \href{http://arxiv.org/abs/math.PR/9904022}{math.PR/9904022}.

\bibitem{taka:93}
Takasaki K., Quasi-classical limit of {BKP} hierarchy and {$W$}-inf\/inity
  symmetries, \href{http://dx.doi.org/10.1007/BF00745149}{\textit{Lett. Math. Phys.}} \textbf{28} (1993), 177--185,
  \href{http://arxiv.org/abs/hep-th/9301090}{hep-th/9301090}.

\bibitem{taka:06}
Takasaki K., Dispersionless {H}irota equations of two-component {BKP}
  hierarchy, \href{http://dx.doi.org/10.3842/SIGMA.2006.057}{\textit{SIGMA}} \textbf{2} (2006), 057, 22~pages,
  \href{http://arxiv.org/abs/nlin.SI/0601063}{nlin.SI/0601063}.


\bibitem{tak-tak:95}
Takasaki K., Takebe T., Integrable hierarchies and dispersionless limit,
  \href{http://dx.doi.org/10.1142/S0129055X9500030X}{\textit{Rev. Math. Phys.}} \textbf{7} (1995), 743--808,
  \href{http://arxiv.org/abs/hep-th/9405096}{hep-th/9405096}.

\bibitem{tak-tak:06}
Takasaki K., Takebe T., Radial L\"owner equation and dispersionless cmKP
  hierarchy, \href{http://arxiv.org/abs/nlin.SI/0601063}{nlin.SI/0601063}.

\bibitem{tak-tak:08}
Takasaki K., Takebe T., L\"owner equations, {H}irota equations and reductions
  of the universal {W}hitham hierarchy, \href{http://dx.doi.org/10.1088/1751-8113/41/47/475206}{\textit{J.~Phys.~A: Math. Theor.}}
  \textbf{41} (2008), 475206, 27~pages, \href{http://arxiv.org/abs/0808.1444}{arXiv:0808.1444}.

\bibitem{t-t-z:06}
Takebe T., Teo L.-P., Zabrodin A., L\"owner equations and dispersionless
  hierarchies, \href{http://dx.doi.org/10.1088/0305-4470/39/37/010}{\textit{J.~Phys.~A: Math. Gen.}} \textbf{39} (2006),
  11479--11501, \href{http://arxiv.org/abs/math.CV/0605161}{math.CV/0605161}.

\bibitem{teo:03}
Teo L.-P., Analytic functions and integrable hierarchies~-- characterization of
  tau functions, \href{http://dx.doi.org/10.1023/A:1024969729259}{\textit{Lett. Math. Phys.}} \textbf{64} (2003), 75--92,
  \href{http://arxiv.org/abs/hep-th/0305005}{hep-th/0305005}.

\bibitem{tsa:90}
Tsar{\"e}v S.P., The geometry of {H}amiltonian systems of hydrodynamic type.
  {T}he generalized hodograph method, \href{http://dx.doi.org/10.1070/IM1991v037n02ABEH002069}{\textit{Math. USSR-Izv.}} \textbf{37}
  (1991), 397--419.

\end{thebibliography}
\end{document}